\documentclass[sigconf]{acmart}

\usepackage{multirow}
\usepackage{yfonts}
\usepackage{longtable}
\usepackage{arydshln}
\usepackage{makecell}
\usepackage{balance}

\usepackage{subcaption}

\copyrightyear{2023}
\acmYear{2023}
\setcopyright{acmlicensed}\acmConference[ICTIR '23]{Proceedings of the 2023 ACM SIGIR International Conference on the Theory of Information Retrieval}{July 23, 2023}{Taipei, Taiwan}
\acmBooktitle{Proceedings of the 2023 ACM SIGIR International Conference on the Theory of Information Retrieval (ICTIR '23), July 23, 2023, Taipei, Taiwan}
\acmPrice{15.00}
\acmDOI{10.1145/3578337.3605122}
\acmISBN{979-8-4007-0073-6/23/07}

\begin{document}

\title{Causal Collaborative Filtering}

\author{Shuyuan Xu}
 \affiliation{
  \institution{Rutgers University}
  \institution{New Brunswick, NJ, US}
    }
 \email{shuyuan.xu@rutgers.edu}

\author{Yingqiang Ge}
 \affiliation{
  \institution{Rutgers University}
  \institution{New Brunswick, NJ, US}
    }
 \email{yingqiang.ge@rutgers.edu}
 
\author{Yunqi Li}
 \affiliation{
  \institution{Rutgers University}
  \institution{New Brunswick, NJ, US}
    }
 \email{yunqi.li@rutgers.edu}

\author{Zuohui Fu}
 \affiliation{
  \institution{Rutgers University}
  \institution{New Brunswick, NJ, US}
    }
 \email{zuohui.fu@rutgers.edu}

\author{Xu Chen}
 \affiliation{
  \institution{Renmin University of China}
  \institution{Beijing, China}
    }
 \email{xu.chen@ruc.edu.cn}

\author{Yongfeng Zhang}
 \affiliation{
  \institution{Rutgers University}
  \institution{New Brunswick, NJ, US}
    }
 \email{yongfeng.zhang@rutgers.edu}
 
\renewcommand{\shortauthors}{Shuyuan Xu et al.}

\ccsdesc[500]{Information systems~Recommender systems}
\ccsdesc[500]{Computing methodologies~Machine learning}

\begin{abstract}
Many of the traditional recommendation algorithms are designed based on the fundamental idea of mining or learning correlative patterns from data to estimate the user-item correlative preference.
However, pure correlative learning may lead to Simpson's paradox in predictions, and thus results in sacrificed recommendation performance. Simpson's paradox is a well-known statistical phenomenon, which causes confusions in statistical conclusions and ignoring the paradox may result in inaccurate decisions. 
Fortunately, causal and counterfactual modeling can help us to think outside of the observational data for user modeling and personalization so as to tackle such issues. 
In this paper, we propose \textbf{C}ausal \textbf{C}ollaborative \textbf{F}iltering (CCF) --- a general framework for modeling causality in collaborative filtering and recommendation. 
We provide a unified causal view of CF and mathematically show that many of the traditional CF algorithms are actually special cases of CCF under simplified causal graphs. We then propose a conditional intervention approach for $do$-operations so that we can estimate the user-item causal preference based on the observational data. Finally, we further propose a general counterfactual constrained learning framework for estimating the user-item preferences. Experiments are conducted on two types of real-world datasets---traditional and randomized trial data---and results show that our framework can improve the recommendation performance and reduce the Simpson's paradox problem of many CF algorithms.

\end{abstract}

\keywords{Collaborative Filtering; Causal Analysis; Counterfactual Reasoning; Simpson's Paradox; Recommender Systems}

\maketitle

\section{Introduction}

Recommender systems are important and valuable tools to provide sophisticated services for many Web-based services such as e-commerce, social networks and online media systems.
Collaborative Filtering (CF) \cite{ekstrand2011collaborative,schafer2007collaborative} algorithms, among others, are fundamental algorithms that support the underlying mechanism of recommender systems.

Most of the existing CF models are developed based on associative user-item preference learning. However, associative learning may be vulnerable to the Simpson's paradox \cite{pearl2016causal,jadidinejad2021simpson} and thus leads to sacrificed recommendation performance.
Simpson's paradox refers to the phenomenon that the statistical conclusion from the total observational data disagrees with that from the sub-groups of data \cite{jadidinejad2021simpson}.
Take Table \ref{tab:paradox} as a toy example, for two candidate items $v_1$ and $v_2$, we have the feedback (e.g., like or dislike) of the users who interacted with $v_1$ or $v_2$, assuming that there are 100 users for each item. Suppose the 100 users can be divided into two groups $G_1$ and $G_2$, e.g., based on gender, age or income. For each group, we have the percentage of users in the group who like the corresponding item, and we can also calculate the overall percentage of users who like each item. 
We can see that it is possible that $v_2$ is more likely to be recommended than $v_1$ according to the data of each group, but $v_1$ is more likely to be recommended than $v_2$ according to the overall data, leading to the Simpson's paradox.

\begin{figure}[t!]
\captionsetup[sub]{font=small,labelfont=normalfont,textfont=normalfont}
    \centering
    \begin{subfigure}[t]{0.22\textwidth}
        \includegraphics[scale=0.35]{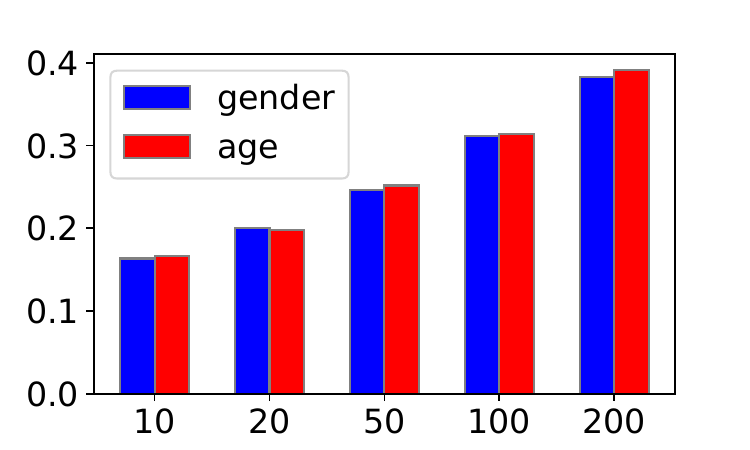}
        \vspace{-15pt}
        \caption{}
    \end{subfigure}
    \begin{subfigure}[t]{0.22\textwidth}
        \includegraphics[scale=0.35]{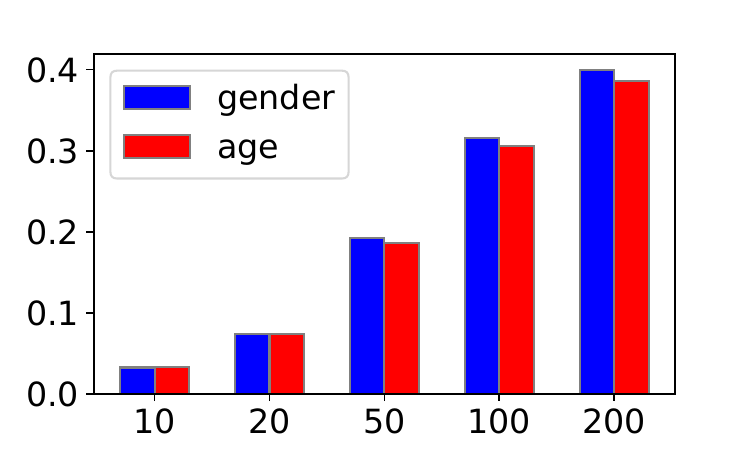}
        \vspace{-15pt}
        \caption{}
    \end{subfigure}
    \vspace{-13pt}
    \caption{(a) The percentage of item pairs that have paradox in the observational data of MovieLens-100k. (b) The percentage of item pairs that have paradox on the full user-item ranking score matrix completed by MF on MovieLens-100K.}
    \label{fig:paradox}
    \vspace{-20pt}
\end{figure}

Such Simpson's paradox also exists for real-world data. 
Following the paradox detection method in \cite{jadidinejad2021simpson}, we show observations on the MovieLens-100K data. 
For each user, we rank the user's interacted items according to ratings and each user's top-$K$ items are considered recommended by the user.
Figure \ref{fig:paradox}(a) shows the percentage of item pairs that have paradox among all possible item pairs in the observational data, with $K$ ranging from 10 to 200 and users grouped by gender or age (age threshold is 35). We can see that a large percentage of item pairs results in Simpson's paradox.

Additionally, such paradox could be learned into associative CF models which leads to paradoxes in the final recommendation list.
Take Matrix Factorization (MF) \cite{rendle2012bpr} on MovieLens-100K as an example. Based on the full user-item ranking score matrix completed by the well-trained MF model, each user recommends his or her top-$K$ ranked items and for each item we randomly sample 100 users.
Figure \ref{fig:paradox}(b) shows the percentage of item pairs that have paradox with $K$ ranging from 10 to 200, which indicates the existence of Simpson's paradox in the final recommendation lists.
As a result, Simpson's paradox exists in both observational data and the predicted data, which may mislead the recommendation results. 
Thus mitigating Simpson's paradox will help improve the recommendation performance (we will show that in Section \ref{sec:experiments}).

One important approach to mitigating Simpson's paradox is causal inference and $do$-operations \cite{pearl2016causal}. 
In this paper, we propose a causal collaborative filtering (CCF) model as a simple and principled framework that seamlessly integrates causal inference and recommendation for reduced paradox and better recommendations. 

Intuitively, estimating the user-item causal preference can be interpreted as answering a \textit{what if} question: what would be the user's preference on an item if we intervene to recommend the item to the user \cite{wang2020causal}. Using standard mathematical language of causal inference \cite{pearl2016causal}, the above \textit{what if} question can be represented as $P(y|u,do(v))$, 
where $u,v$ is a user-item pair and $y$ is the preference score to be estimated for the pair, e.g., $y=1$ for likes and $0$ for dislikes. In the CCF framework, $do$-operation is used to represent the causal preference if we intervene to recommend item $v$ instead of passively observing item $v$ in training data.
More interestingly, we show that traditional CF models are actually special cases of CCF under simplified causal graphs (Figure \ref{fig:causal_graph}), and CCF is a general framework for casual learning in recommendation which can be applied over various causal graphs.

Except for the above conceptual contribution, this work also provides technical contributions. More specifically, a great challenge is how to estimate $P(y|u,do(v))$. 
In this work, we propose a conditional intervention approach to estimating $P(y|u,do(v))$ based on observational data. 
Specifically, we adopt the causal graph in Figure \ref{fig:causal_graph}(d) for conditional intervention, which considers the user interaction history $X$ for mediator analysis. Moreover, solving the conditional intervention requires counterfactual reasoning, and we propose a counterfactual constrained learning framework for counterfactual reasoning in both discrete and continuous space to estimate the causal preference $P(y|u,do(v))$.
We conduct extensive experiments on real-world datasets. Experimental results show that CCF reduces Simpson's paradox and significantly improves the recommendation performance.

\begin{table}[t]
    \centering
    \small
    \begin{tabular}{ccc}
    \toprule
         & Item $v_1$ & Item $v_2$ \\\midrule
        User Group $G_1$ & 62.5\% (50/80) & \textbf{66.6\%} (20/30)\\
        User Group $G_2$ & 50.0\% (10/20) & \textbf{54.3\%} (38/70)\\\midrule
        Overall & \textbf{60.0\%} (60/100) & 58.0\% (58/100)\\
    \bottomrule
    \end{tabular}
    \caption{A toy example of Simpson's paradox in recommendation, where two candidate items $v_1$ and $v_2$ are considered. ($x/y$) represents that there are $x$ users like the item within $y$ users who have interacted with the item.
    }
    \vspace{-20pt}
    \label{tab:paradox}
    
\end{table}

\section{Related Work}\label{sec:related}

Existing literature usually categorizes the recommendation algorithms into three major types: collaborative filtering, content-based recommendation and hybrid method \cite{jannach2010recommender, zhang2019deep, adomavicius2005toward}. Due to the wide scope of literature of recommender systems (RS), it is hardly possible to cover all of the RS algorithms, so we review some representative methods based on collaborative filtering in this section, and a more comprehensive review can be seen in \cite{ekstrand2011collaborative,zhang2019deep,jannach2010recommender, adomavicius2005toward}.

Collaborative Filtering (CF) \cite{ekstrand2011collaborative} is based on a key idea that similar users may share similar interests and similar items may be liked by similar users.
Early memory-based CF models---such as user-based CF \cite{resnick1994grouplens,konstan1997grouplens} and item-based CF \cite{sarwar2001item,linden2003amazon}---calculate the similarity between users or items for recommendation based on pre-defined similarity functions such as cosine similarity. To extract latent semantic meanings from the matrix, researchers later explored learned user and item vector representations to calculate the matching score of each user-item pair for recommendation, including Latent Factor Models (LFM) such as matrix factorization \cite{koren2009matrix}, tensor factorization \cite{karatzoglou2010multiverse} and factorization machines \cite{rendle2010factorization}. 
The development of deep learning and neural networks has further extended CF. The relevant methods can be broadly classified into two categories: similarity learning approach and representation learning approach. The similarity learning approach adopts simple user and item representations (such as one-hot) and learns a complex matching function (such as a prediction network) to calculate user-item matching scores \cite{cheng2016wide,xue2017deep,hsieh2017collaborative}, while the representation learning approach learns rich user and item representations and adopts a simple matching function (e.g., inner product) for efficient matching score calculation \cite{zheng2017joint,zhang2017joint,zhang2016collaborative,ai2018learning,mcauley2015image}. Another important direction is learning to rank for recommendation, which learns the relative ordering of items instead of the absolute scores, such as Bayesian Personalized Ranking (BPR) \cite{rendle2012bpr}.

Most existing methods learn correlative patterns from data for matching and recommendation based on either simple or complex matching functions. However, advancing from correlative learning to causal learning is an important problem \cite{pearl2016causal}. The community has explored causal modeling on several different perspectives.
For example, researchers adopted causal models to generate explanations for recommendation \cite{ghazimatin2020prince,tran2021counterfactual,tan2021counterfactual}, considered fairness under counterfactual settings \cite{li2021towards,mehrotra2018towards}, corrected data bias for rankings in search \cite{joachims2017unbiased,ai2018unbiased,oosterhuis2020unbiased,wang2018position,hu2019unbiased,oosterhuis2020policy}, recommendation \cite{bonner2018causal,schnabel2016recommendations,liu2020general,wang2020causal,sato2020unbiased,zheng2021disentangling,zhang2021causal,wang2021deconfounded,wu2021unbiased,wei2021model}, advertising \cite{yuan2019improving} and evaluating the ranking models \cite{yang2018unbiased}, estimated the uplift effect of recommendations \cite{sato2019uplift,sato2020unbiased,xie2021causcf}, explored data augmentation \cite{wang2021counterfactual,yang2021top,zhang2021causerec} as well as multimodal information such as text \cite{xiong2021counterfactual} based on causal methods. 
Some related works are proposed to estimate $P(y|u,do(v))$ for recommendation as well, for example, \citet{zhang2021causal} leverage popularity bias for recommendation, \citet{xu2022dynamic} design a causal model for mitigating echo chambers while maintaining comparable performance, etc. Unlike the existing causal recommendation works, our work focuses on mitigating Simpson's paradox. 

Simpson's paradox is a common statistical phenomenon and it appears in many artificial intelligence applications and real-life scenarios \cite{neufeld1995simpson,von2021simpson}. This phenomenon was originally observed in 1951 \cite{simpson1951interpretation} and was later named as the ``Simpson's Paradox''\cite{blyth1972simpson}. Simpson's paradox has attracted the attention of many computer scientists in recent years. In general machine learning, existing literature mainly focuses on detecting Simpson' paradox automatically \cite{alipourfard2018can,alipourfard2018using,xu2018detecting,sharma2022detecting,shmueli2018forest,sharma2022detecting}. In recommender systems, \citet{jadidinejad2021simpson} propose a method to address the Simpson's paradox in offline evaluation.
To the best of our knowledge, none of the existing works aims at proposing models to mitigate Simpson's paradox in model prediction for improved recommendation performance.

\section{A Unified Causal View of CF}
\label{sec:unified}
We provide a unified causal view of collaborative filtering (CF) in this section. Specifically, we show that the fundamental goal of many CF algorithms is to estimate the causal effect $P(y|u,do(v))$. 
The key difference between various CF models is that they assume different causal graphs to calculate $P(y|u,do(v))$. When the causal graph is too simple or even unrealistic, the causal effect will naturally degenerate to association relations that are considered in traditional CF models.
We now show how different CF models fit into the unified causal view under $P(y|u,do(v))$.

\subsection{Non-Personalized Model}
Non-personalized recommendation models, such as most popular recommendation \cite{ji2020re}, assume a simple causal graph without the user node, as shown in Figure \ref{fig:causal_graph}(a). Since user is excluded from consideration and since item is a root node in the graph, we have $P(y|u,do(v))=P(y|do(v))=P(y|v)$, and $P(y|v)$ naturally represents the popularity of item $v$ in the data.

\vspace{-1ex}
\subsection{Associative Matching Models}
Most CF algorithms fall into the user-item associative matching category.
These models assume a causal graph shown in Figure \ref{fig:causal_graph}(b), where user node $U$ and item node $V$ constitute a collider to influence preference node $Y$. 
Basically, these models assume that the appearance of users and items are independent from each other in observational data (though this may be an unrealistic assumption), and since both $U$ and $V$ are root nodes, we have $P(y|u,do(v))=P(y|u,v)$, which can thus be estimated from observational data.

The main difference of various models is how to design the matching function to estimate $P(y|u,v)$, e.g., user-based CF assumes $P(Y=1|u,v)\propto\frac{1}{|N(u)|}\sum_{u^\prime\in N(u)}y_{u^\prime,v}$, where $N(u)$ are the neighbours of user $u$.
Matrix factorization (MF) models, such as \cite{koren2009matrix}, assume $P(Y=1|u,v)\propto\mathbf{u}^\intercal\mathbf{v}$ or $\propto\mathbf{u}^\intercal\mathbf{v}+b_u+b_v+b$. 
Some neural network-based models such as \cite{he2017neural,xue2017deep,cheng2016wide} assume $P(Y=1|u,v)\propto \text{NN}(\mathbf{u},\mathbf{v})$, where NN is a neural network for similarity matching.
More complex deep representation learning models such as sequential models \cite{hidasi2016session,chen2018sequential,li2017neural,tang2018personalized} and graph-based models \cite{zhang2016collaborative,ai2018learning,ying2018graph,wang2019unified} can be represented as $P(Y=1|u,v)\propto\text{NN}(\textbf{NN}(u),\textbf{NN}(v))$, where a neural similarity network NN is applied on top of the neural representation learning network \textbf{NN}.

\vspace{-1ex}
\subsection{Causal Reasoning Model}
Some causal models \cite{bonner2018causal,schnabel2016recommendations,joachims2018deep} are aware of the dependencies between user and item, thus assume the causal graph in Figure \ref{fig:causal_graph}(c), which extends Figure \ref{fig:causal_graph}(b) by removing the independence assumption between user and item.
In this case, the $u$-specific causal effect $P(y|u,do(v))$ by definition requires interventional reasoning. Depending on if or not we have complete control of the recommendation platform, we have the following two approaches to estimate $P(y|u,do(v))$.

\subsubsection{\textbf{Direct Intervention Models}}
\begin{figure}[t]
\vspace{-10pt}
    \centering
    \includegraphics[width=\linewidth]{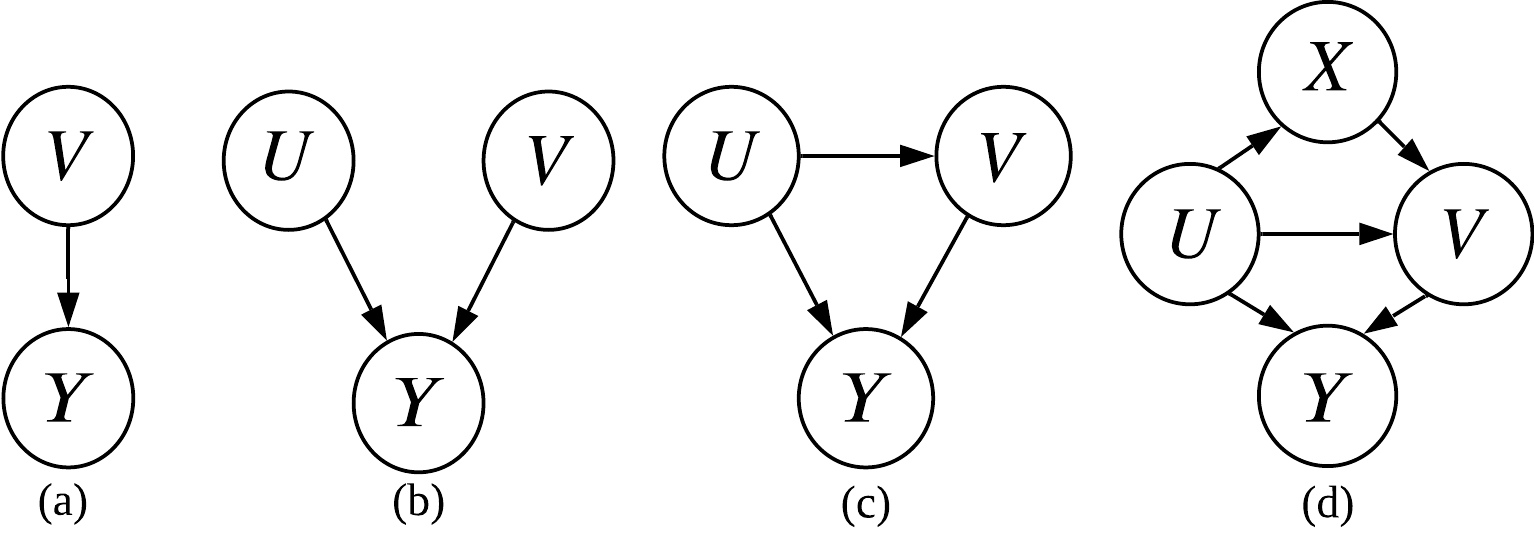}
    \vspace{-20pt}
    \caption{Many traditional CF models are special cases of CCF under simplified causal graphs. In the graphs, $U$ is user, $V$ is item, $X$ is user interaction history, $Y$ is preference score. (a) Causal graph for non-personalized models. (b) Causal graph for similarity matching-based CF models. (c) Causal graph that considers the causality from user to item \cite{bonner2018causal}. (d) Causal graph used in our framework to demonstrate the idea of CCF, using user interaction history $X$ as a mediator.
    }
    \label{fig:causal_graph}
    \vspace{-10pt}
\end{figure}

If we have complete control of the recommendation platform or have access to a randomized treatment dataset where user is randomly exposed to items, then the straightforward way of estimating $P(y|u,do(v))$ is through direct intervention \cite{bonner2018causal,wang2020causal,zhang2021causal}. 
We refine Figure \ref{fig:causal_graph}(c) as Figure \ref{fig:causal}(a) to show the structural equations $V=g(U)$ and $Y=f(U,V)$, which represent the two steps of the recommendation pipeline.
$V=g(U)$ represents the de facto recommendation model in the system that decides what items are exposed to the user, and $Y=f(U,V)$ represents the user's preference on the exposed item. To estimate $P(y|u,do(v))$, we resort to the most original definition of intervention to get the manipulated causal graph as shown in Figure \ref{fig:causal}(b) \cite[p.54]{pearl2016causal}.
We thus have $P(y|u,do(v))=P_m(y|u,v)$, where $P_m$ is the probability distribution according to the manipulated causal graph. To estimate $P_m(y|u,v)$, we can apply a randomized exposure policy by either showing random items to users or manipulating the observational data to simulate a random policy and thus to implement the independence between $U$ and $V$. This treatment will help us to collect an unbiased dataset to estimate $P_m(y|u,v)$. More details can be seen in \cite{bonner2018causal,wang2020causal,zhang2021causal}.

\vspace{-5pt}
\subsubsection{\textbf{Inverse Propensity Scoring (IPS) Models}} 
In many cases, we do not have complete control of the recommendation platform or access to the randomized treatment data. 
The basic idea of inverse propensity scoring (IPS) methods is to turn the outcomes of an observational study into pseudo-randomized trials by re-weighting the samples \cite{bonner2018causal}, so that $P(y|u,do(v))$ can be estimated from the observational data \cite{schnabel2016recommendations,joachims2018deep}.
More formally, according to the recommendation pipeline shown in Figure \ref{fig:causal}(a), the observed user preference $r_{uv}$ is considered as $r_{uv}\propto P(y|u,do(v))P(v|u)$, which is the multiplication between the user's real preference $P(y|u,do(v))$ and the probability that user $u$ had a chance to see the item $P(v|u)$. As a result, we have $P(y|u,do(v))\propto \frac{r_{uv}}{P(v|u)}$, which means that each example in the observational data boosts its probability by a factor equal to $1/P(v|u)$,
which corrects the observational data by removing the exposure bias.

\vspace{-1ex}
\section{The Proposed Framework}
\label{sec:ccf_model}

In this section, we will start from the causal graph and then introduce the techniques for estimating $P(y|u,do(v))$, including conditional intervention, counterfactual reasoning, and finally a flexible counterfactual constrained learning framework that can be applied on any existing CF model for recommendation.

\vspace{-5pt}
\subsection{The Causal Graph}
As mentioned above, in many cases of recommender systems we want to answer counterfactual questions
such as \textit{what if} an item had (or had not) been recommended, 
or \textit{what if} the user had a different interaction history. Such imaginary cases constitute the \textit{counterfactual world}, in contrast to what happened in the \textit{real world}.

\begin{figure}[t]
\vspace{-5pt}
    \centering
    \includegraphics[width=\linewidth]{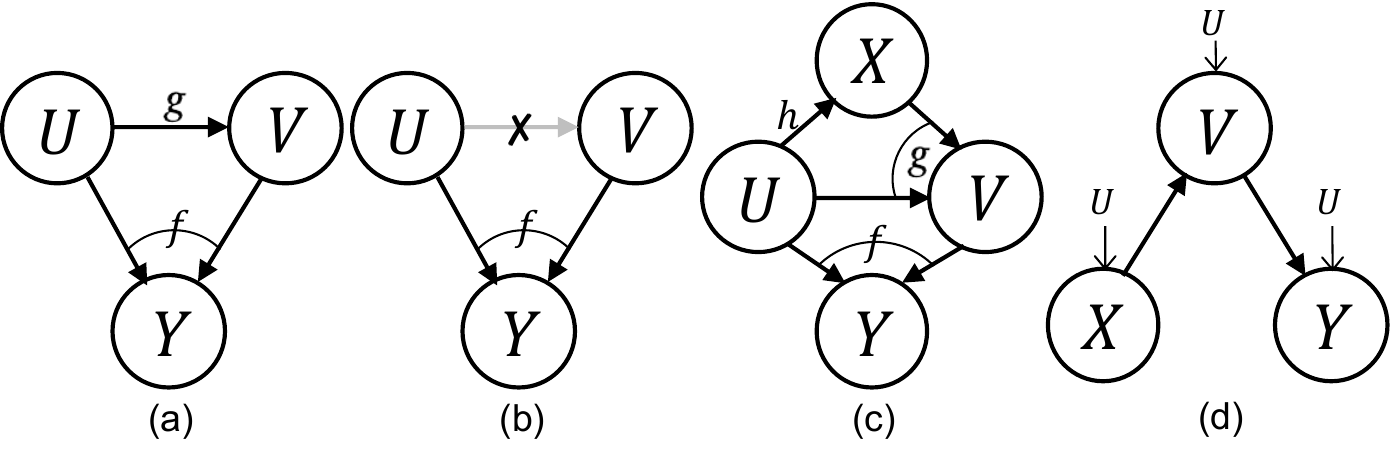}
    \vspace{-27pt}
    \caption{(a-b) Causal graphs before and after manipulation. (c-d) Reorganize causal graph using $U$ as exogenous variable.}
    \label{fig:causal}
    \vspace{-15pt}
\end{figure}

To enable counterfactual reasoning, we extend the causal graph from Figure \ref{fig:causal}(a) to Figure \ref{fig:causal}(c) to consider user's interaction histories $X$ for mediator analysis. Specifically, the casual model includes three structural equations: (1) $X=h(U)$, which returns a user's history $X$. In the most simple case, it can be a database retrieval operation that returns a user's interaction history;
(2) $V=g(U,X)$, which is the already deployed recommendation algorithm of the system that returns the recommended item $V$ based on the user and the user's interaction history; (3) $Y=f(U,V)$, which is the user preference function that we do not know but we want to estimate. 

We should acknowledge that the causal graph in Figure \ref{fig:causal}(c) is not a once-and-for-all solution for recommender systems, because practical systems are very complicated that involve many other factors.
However, we consider this causal graph in the work because the structural equation $V=g(U,X)$ is general enough to include a wide scope of recommendation algorithms, including both sequential and non-sequential methods.  
With the help of the causal graph, our framework aims to estimate $P(y|u,do(v))$ for reduced paradox and enhanced performance, which we will show in the following.

\vspace{-5pt}
\subsection{Conditional Intervention}
\label{sec:conditional_intervention}
To estimate $P(y|u,do(v))$, we first identify that $\{U,X\}$ is a set of variables that satisfy the backdoor criterion \cite[p.61]{pearl2016causal} for the casual effect $V\rightarrow Y$. Since we already conditioned on $U$ for personalization, the only variable that leads to variations in $V$ is user interaction history $X$, as a result, we adopt conditional intervention \cite[p.70]{pearl2016causal}\cite[p.113]{pearl2000models} to estimate $P(y|u,do(v))$. 

More specifically, the recommendation policy $V=g(U,X)$ provides recommendation $V$ based on the user $U$ and history $X$, written as $do(V=g(U,X))$. To find out the distribution of the outcome $Y$ that results from this policy, we seek to estimate $P(Y=y|U=u,do(V=g(U,X)))$. We will show that identifying the effect of such policies is equivalent to identifying the expression for the $(u,x)$-specific effect $P(Y=y|U=u,X=x,do(V=v))$ \cite[p.71]{pearl2016causal}.
\begin{equation}
\label{eq:conditional_intervention}
\begin{aligned}
    P(y|&u,do(v))\doteq P(Y=y|U=u,do(V=g(U,X)))\\
    \overset{1}{=}\sum_x &P(Y=y|U=u,do(V=g(U,X)),X=x)~\times\\
    &P(X=x|U=u,do(V=g(U,X)))\\
    \overset{2}{=}\sum_x &P(Y=y|U=u,X=x,do(V=g(u,x)))P(X=x|U=u)\\
    \overset{3}{=}\sum_x &P(Y=y|U=u,X=x,do(V=v))|_{v=g(u,x)} P(X=x|U=u)\\
    \overset{4}{=}\sum_x &P(y|u,x,v)|_{v=g(u,x)} P(x|u)=E_{x|u}[P(y|u,x,v)|_{v=g(u,x)}]
\end{aligned}
\end{equation}

From the last step in Eq.\eqref{eq:conditional_intervention} we can see that the key difference between the causal model $P(y|u,do(v))$ and traditional associative models $P(y|u,v)$ is the existence of the conditional probability term $P(x|u)$ in the final step. In step 4, $P(y|u,x,v)|_{v=g(u,x)}$ stands for the preference estimation of the deployed recommendation model $V=g(U,X)$. 
Traditional models only consider the real world but not the counterfactual world, as a result, the conditional probability $P(x|u)=1$ for observed user history $x$, while for unobserved history $x^\prime$, $P(x^\prime|u)=0$. In this case, we see that the summation in step 4 will only include observed history $x$ and thus $P(y|u,do(v))$ naturally degenerates to the original recommendation model $V=g(U,X)$.

However, the observed history $x$ does not mean that the user is destined to interact with the items in $x$---the user just happened to interact with $x$, i.e., if the user had a chance to be recommended with different items $x^\prime$ in the counterfactual world, the user may also interact with those items, and thus the probability $P(x^\prime|u)$ is not 0. As a result, the calculation of Eq.\eqref{eq:conditional_intervention} requires counterfactual reasoning where the user history had been $X=x^\prime$, which is beyond the observational data $X=x$.

\begin{table*}[t]
    \centering
    \caption{Different heuristic rules to create counterfactual examples, the corresponding counterfactual question, and some intuitive toy examples. In the toy examples, the user's real interaction history $x$ includes items $a~b~c$, and items at the right side of the arrow is the counterfactual history $x^\prime$. Multiple counterfactual histories can be constructed from the real history $x$.}
    \vspace{-10pt}
    \begin{tabular}{p{2.2cm}|p{8cm}|p{6cm}}
    \toprule
        Heuristic Rule & Counterfactual Question & Toy Example\\
        \midrule
        Keep One (K1) & What if the user only interacted with one history item? & $a~b~c\rightarrow a; ~a~b~c\rightarrow b; ~a~b~c\rightarrow c$\\
        Delete One (D1) & What if the user did not interact with one of the history items? & $a~b~c\rightarrow b~c;~a~b~c\rightarrow a~c;~a~b~c\rightarrow a~b$\\
        Replace One (R1) & What if one of the history items were different? & $a~b~c\rightarrow a^\prime b~c;~a~b~c\rightarrow a~b^\prime c;~a~b~c\rightarrow a~b~c^\prime$\\\bottomrule
        
    \end{tabular}
    \label{tab:heuristic_rules}
    \vspace{-10pt}
\end{table*}

\vspace{-1ex}
\subsection{Counterfactual Reasoning}
\label{sec:counterfactual_reasoning}
Counterfactual reasoning enables more refined intervention at individual level \cite[p.78,93]{pearl2016causal}. In this work, the \textit{individual level} refers to each user $U=u$ for personalization purpose. To better understand this, the causal graph in Figure \ref{fig:causal}(c) is equivalently transformed into Figure \ref{fig:causal}(d), where $U$ serves as the exogenous variable. As a result, counterfactual reasoning is individualized on each user.

To enable counterfactual reasoning to calculate Eq.\eqref{eq:conditional_intervention}, let's consider a record $(u,x,v,y)$ in the observational data, meaning that user $u$'s real history is $x$, and then the system logged user's preference on item $v$ which is $y$, e.g., we can consider binary preference values using $y=1$ for likes and $y=0$ for dislikes, but the framework can also be applied over multiple preference values.
According to Eq.\eqref{eq:conditional_intervention}, the user preference estimation $y=f(u,v)$ is expressed as
\begin{equation}
\hspace{-8pt}
\begin{aligned}
\label{eq:unbiased}
y&=f(u,v)\propto P(y|u,do(v))\\
&=\sum_{\tilde{x}} P(y|u,\tilde{x},v)|_{v=g(u,\tilde{x})} P(\tilde{x}|u)=E_{\tilde{x}|u}[P(y|u,\tilde{x},v)|_{v=g(u,\tilde{x})}]
\end{aligned}
\end{equation}
To distinguish from the single real-world history $x$, we use $\tilde{x}$ to represent any possible user history, including both the real history $x$ and possible counterfactual histories $x^\prime$. Eq.\eqref{eq:unbiased} means that the estimation of $P(y|u,do(v))$ can be achieved by correcting the original recommendation algorithm's estimation $P(y|u,x,v)|_{v=g(u,x)}$ using counterfactual histories $x^\prime$. More specifically, the estimation for $P(y|u,do(v))$ is the \textit{expected} estimation of $P(y|u,x,v)|_{v=g(u,x)}$, where the expectation is taken over all possible histories (including real and counterfactual histories) when item $v$ is recommended.

\subsubsection{\textbf{Generate Counterfactual Examples}}
\label{sec:generate}
Counterfactual reasoning requires generating counterfactual examples based on minimal changes \cite[p.92]{pearl2016causal}. 
We start with a heuristic-based approach for counterfactual example generation and we will generalize to a learning-based approach in the next section.

We adopt three heuristic rules to generate counterfactual histories $x^\prime$ by applying modifications to the real history $x$ (Table \ref{tab:heuristic_rules}). The Keep One (K1) rule only keeps one item of the user's real history, the Delete One (D1) rule removes one item from the user's real history, 
and the Replace One (R1) rule replaces one item of the user's real history with another item. For the R1 rule, depending on how the item is replaced, we have two variants: R1-random (R1r)---the item is replaced with a random item, and R1-nearest (R1n)---the item is replaced with its nearest neighbour based on embedding similarity. We will introduce more details in the experiments. 

\subsubsection{\textbf{Select Counterfactual Examples}}
\label{sec:select} 
Consider the training example $(u,x,v,y)$ where the user's real history is $x$, and we are able to generate $m$ counterfactual histories $\{x_1^\prime, x_2^\prime \cdots x_m^\prime\}$ using one of the heuristic rules.
Conditional intervention (Section \ref{sec:conditional_intervention}) requires $v=g(u,\tilde{x})$, i.e., the same item $v$ should be recommended (i.e., within the top-$k$ recommendation list) by the recommendation algorithm $g(\cdot,\cdot)$ under counterfactual histories (since we are considering $do(v)$ instead of just $v$ in the condition). However, not all of the counterfactual histories $\{x_1^\prime, x_2^\prime \cdots x_m^\prime\}$ guarantee that item $v$ is recommended under the algorithm. As a result, we execute the recommendation algorithm $g(\cdot,\cdot)$ over each counterfactual history $x_i^\prime$ and obtain the top-$k$ recommendation list $\mathcal{V}_i^\prime=g(u,x_i^\prime)$, where $k$ is a hyper-parameter to be tuned (will be introduced in the experiments). If the target item $v\in\mathcal{V}_i^\prime$, then we keep the counterfactual example $(u,x_i^\prime,v,y)$.
Suppose $n$ of the $m$ counterfactual histories are eventually selected, we will have a set of counterfactual examples $\{(u,x_i^\prime,v,y)\}_{i=1}^{n}$.

\subsubsection{\textbf{Calculate the Expectation}}
We then calculate $P(y|u,do(v))$ based on the real observation $(u,x,v,y)$ and the counterfactual examples $\{(u,x_i^\prime,v,y)\}_{i=1}^{n}$ according to Eq.\eqref{eq:unbiased}. For simplicity, we consider $P(\tilde{x}|u)$ as a piecewise uniform distribution over the real and counterfactual histories, i.e., 
\begin{equation}\label{eq:uniform}
P(\tilde{x}|u) = 
\begin{cases} 
\alpha,~ \text{when}~\tilde{x} = x \\ 
\beta,~ \text{when}~\tilde{x} = x_i^\prime,~ i\in\{1,2\cdots n\}
\end{cases},
\alpha+n\beta=1
\end{equation}
where $\alpha$ is the probability of the real example $x$, and $\beta$ is the probability of each counterfactual example $x_i^\prime$. Since $x$ is already observed, we apply a higher probability to $x$ than $x_i^\prime$, i.e., $\alpha>\beta>0$.
Generalizing to even more complex distributions such as Gaussian or Gamma distribution will be considered in the future. Then we have:
\begin{equation}
\begin{aligned}
\label{eq:expectation}
    P(y|u,do(v)) &= \sum_{\tilde{x}} P(y|u,\tilde{x},v)|_{v=g(u,\tilde{x})} P(\tilde{x}|u)\\ 
    &= \alpha~ P_g(y|u,x,v)+\beta \sum_{i=1}^n P_g(y|u,x_i^\prime,v)
\end{aligned}
\end{equation}
where $P_g$ is the probability estimation of the base recommendation algorithm $v=g(u,x)$.

\subsection{Counterfactual Constrained Learning}
In practical recommender systems, the ranking probability score $P_g(y|u,x,v)$ is usually learned by optimizing a loss function $L(g)$ such as the rating prediction loss \cite{koren2009matrix} or the pair-wise ranking loss \cite{rendle2012bpr}. As noted before, however, the estimated probability $P_g(y|u,x,v)$ could be unreliable due to unrealistic model assumptions or data bias. 
As a result, what we really want is the probability score of $P(y|u,do(v))$ for item ranking. To learn the values of $P(y|u,do(v))$, we propose a counterfactual constrained learning approach, which requires the base recommender's probability estimation $P_g(y|u,x,v)$ to be equal to $P(y|u,do(v))$, and thus we can safely use the learned $P_g(y|u,x,v)$ scores for item ranking and recommendation:
\begin{equation}
\begin{aligned}
\label{eq:obj}
    &\text{minimize}~L(g)\\
    \text{s.t.}~ &P_g(y|u,x,v) = P(y|u,do(v)) \quad \forall u\in \mathcal{U},~\forall v \in \mathcal{V}
\end{aligned}
\end{equation}
where $L(g)$ is the loss function of a base recommendation algorithm $g(u,x)$, $\mathcal{U}$ is the set of users, and $\mathcal{V}$ is the set of items.

Actually, the constraint $P_g(y|u,x,v) = P(y|u,do(v))$ is naturally supported by the causal graph (Figure \ref{fig:causal}(c)). The reason is that in the graph, both $U$ alone and $\{U,X\}$ as a set satisfy the backdoor criterion for $V\rightarrow Y$, as a result, we have $P_g(y|u,x,v) = P(y|u,do(v))$. Careful readers may ask if $P_g(y|u,x,v) = P(y|u,do(v))$, then why can't we just directly use the estimation $P_g(y|u,x,v)$ of the base recommender for recommendation? The reason is that we only have the accurate $P_g(y|u,x,v)$ scores for the observed $(u,v)$ pairs in the dataset, which are the already observed user preference (ratings, clicks, etc.) on the item. These pairs do not need any estimation and according to the causal graph they can be used as $P(y|u,do(v))$ because the system already exposed the items to the user and collected the user's preference.
However, recommender system needs the $P_g(y|u,x,v)$ scores for the unobserved $(u,v)$ pairs to make recommendations, and these scores need to be estimated using a model. As discussed before, most traditional CF models assume simplified causal graphs to estimate $P_g(y|u,x,v)$ based on associative learning, which may lead to unreliable or even biased estimations. As a result, we need to explicitly add the constraint to the learning procedure to make sure $P_g(y|u,x,v) = P(y|u,do(v))$ is guaranteed for both observed and unobserved $(u,v)$ pairs. In the following, we further derive Eq.\eqref{eq:obj} to make it learnable.

\subsubsection{\textbf{Counterfactual Learning in Discrete Space}}
We first propose a discrete version of the counterfactual constrained learning algorithm for any base recommender, which conducts counterfactual reasoning in a discrete item space. We already know the expression for $P(y|u,do(v))$ (i.e., Eq.\eqref{eq:expectation}). 
Using $P(y|u,x,v)$ for $P_g(y|u,x,v)$ for notation simplicity, the constraint can be written as:
\begin{equation}
\label{eq:cc_transform}
\begin{aligned}
    P(y|u, &x,v) = P(y|u,do(v)) = \alpha~ P(y|u,x,v)+\beta \sum_{i=1}^n P(y|u,x_i^\prime,v)\\
    \Leftrightarrow ~~~& (1-\alpha)~P(y|u,x,v) = \beta\sum_{i=1}^n P(y|u,x_i^\prime,v)\\
    \Leftrightarrow ~~~& \sum_{i=1}^n P(y|u,x_i^\prime,v)=\frac{1-\alpha}{\beta}\cdot P(y|u,x,v)=n\cdot P(y|u,x,v)\\
    \Leftrightarrow ~~~& \sum_{i=1}^n P(y|u,x_i^\prime,v)-n\cdot P(y|u,x,v)=0
\end{aligned}
\end{equation}

To make the constraint optimizable, we relax the equality constraint to an inequality constraint, i.e.,

\begin{equation}
\begin{aligned}
\label{eq:inequalconstraint}
    &\text{minimize}~L(g)\\
    \text{s.t.}~ &\Big|\sum_{x^\prime\in\mathcal{C}(u,v)} P(y|u,x^\prime,v)-|\mathcal{C}(u,v)|\cdot P(y|u,x,v)\Big|\leq\epsilon \\
    \quad &\forall u\in \mathcal{U},~\forall v \in \mathcal{I}(u)\cup\mathcal{S}(u)
\end{aligned}
\vspace{-5pt}
\end{equation}
where $\mathcal{C}(u,v)$ is the set of counterfactual histories of user $u$ under the target item $v$ (section \ref{sec:generate} and \ref{sec:select}), $|\mathcal{C}(u,v)|$ represents the size of the set (i.e. $n$ in Eq.\eqref{eq:cc_transform}), $\mathcal{I}(u)$ is the set of interacted items of user $u$, and $\epsilon$ is a parameter controlling how rigorous is the constraint. 
In Eq.\eqref{eq:inequalconstraint}, since the item space is very huge, it is impractical to apply the constraint on all items in practice. As a result, we sample a set of items for each user, i.e., $\mathcal{S}(u)$, where $|\mathcal{S}(u)|=|\mathcal{I}(u)|$.
For easy implementation, we apply absolute value inequality to constrain the upper bound of the above inequality: 
\begin{equation}
\begin{aligned}
\label{eq:inequality}
    &\Big|\sum_{x^\prime\in\mathcal{C}(u,v)} P(y|u,x^\prime,v)-|\mathcal{C}(u,v)|\cdot P(y|u,x,v)\Big|\\
    \leq &\sum_{x^\prime\in\mathcal{C}(u,v)}\Big|P(y|u,x^\prime,v)- P(y|u,x,v)\Big|\le\epsilon
\end{aligned}
\vspace{-5pt}
\end{equation}
Therefore, we define the final counterfactual learning in discrete space as following:
\begin{equation}
\begin{aligned}
\label{eq:discrete}
    &\text{minimize}~L(g)\\
    \text{s.t.}~ &\sum_{x^\prime\in\mathcal{C}(u,v)}\Big|P(y|u,x^\prime,v)- P(y|u,x,v)\Big|\leq\epsilon \\
    \quad &\forall u\in \mathcal{U},~\forall v \in \mathcal{I}(u)\cup\mathcal{S}(u)
\end{aligned}
\vspace{-5pt}
\end{equation}

According to Eq.\eqref{eq:inequality}, we can see that satisfying the constraint in Eq.\eqref{eq:discrete} naturally leads to satisfying the constraint in Eq.\eqref{eq:inequalconstraint}.

\subsubsection{\textbf{Counterfactual Learning in Continuous Space}}
Many recommendation models represent users, items and histories as embedding vectors in a latent space. If a user's history $x$ is represented as an embedding vector $\mathbf{x}$, then we can directly create latent counterfactual histories $\mathbf{x}^\prime$ by slightly perturbing vector $\mathbf{x}$ in the latent space.  
Similarly, let $L(g)$ be the loss function of a base recommendation algorithm $g(u,x)$, CCF in continuous space aims to learn $L(g)$ under a continuous counterfactual constraint:
\begin{equation}
\begin{aligned}
\label{eq:continuous}
    &\text{minimize}~L(g)\\
    \text{s.t.}~ &\int_{\mathbf{x}^\prime}\big|P(y|u,x^\prime,v)-P(y|u,x,v)\big|\le\epsilon_1,~\|\mathbf{x}^\prime-\mathbf{x}\|_2\le\epsilon_2\\
    \quad &\forall u\in \mathcal{U},~\forall v \in \mathcal{I}(u)\cup\mathcal{S}(u)
\end{aligned}
\vspace{-5pt}
\end{equation}
where $\mathbf{x}$ is the embedding of user $u$'s real history $x$, $\mathbf{x}^\prime$ is a latent vector selected from the small $\epsilon_2$-neighbourhood of vector $\mathbf{x}$, and the integration can be calculated based on Monte Carlo sampling. All other parameters have the same meaning as Eq.\eqref{eq:discrete}.

\vspace{-5pt}
\subsection{Model Learning and Optimization}
To solve the above constrained optimization problem, we formulate the problem as a tractable optimization problem by the Lagrange Multiplier Method. 
For the discrete space version, we convert the objective in Eq.\eqref{eq:discrete} to the following Lagrange optimization form:
\begin{equation}\label{eq:discrete_learning}
\small
\begin{aligned}
    &\text{minimize}~L(g) + \omega L_c \\ 
    & L_c=\sum_{u\in\mathcal{U}}\sum_{v\in \mathcal{I}(u)\cup\mathcal{S}(u)}\text{max}\Big\{0, \sum_{x^\prime\in\mathcal{C}(u,v)}\big|P(y|u,x^\prime,v)-P(y|u,x,v)\big|-\epsilon\Big\}
\end{aligned}
\end{equation}
where $\omega$ is a parameter controlling the weight of the constraint.

While for continuous space, we process the constraint similarly. The difference is that the parameter $\epsilon_2$ in Eq.\eqref{eq:continuous} is used to restrict the distance between counterfactual histories and the real history. 
\begin{equation}\label{eq:continuous_learning}
\small
    \begin{aligned}
        &\text{minimize}~L(g) + \omega L_c\\
        \text{s.t.}~ &~\|\mathbf{x}^\prime-\mathbf{x}\|_2\le\epsilon_2\\
     &L_c=\sum_{u\in\mathcal{U}}\sum_{v\in \mathcal{I}(u)\cup\mathcal{S}(u)}\text{max}\Big\{0,\int_{\mathbf{x}^\prime}\big|P(y|u,x^\prime,v)-P(y|u,x,v)\big|-\epsilon_1\Big\}
    \end{aligned}
\end{equation}
We still apply Monte Carlo sampling for integration calculation.
The counterfactual constrained learning framework is flexible and can be applied on many base recommender algorithms $g$, which we will show in the experiments.

\section{Experiments}\label{sec:experiments}

We conduct experiments to explore CCF from different perspectives. In particular, we aim to answer the following research questions: \textbf{RQ1}: What is the overall performance of the CCF framework, can CCF improve the recommendation performance? \textbf{RQ2}: Can CCF reduce Simpson's paradox? \textbf{RQ3}: How different heuristic rules influence the performance? \textbf{RQ4}: Is it necessary to select the counterfactual examples after they are generated? \textbf{RQ5}: What is the impact of the counterfactual constraint in the learning objective?
We will first describe the datasets, baselines and then provide our answers to the above questions.

\begin{table*}[t]
    \footnotesize
\centering
\setlength{\tabcolsep}{3pt}
\caption{Performance of all three recommendation models on MovieLens-100k and Coat Shopping. The relative improvement is calculated against the original performance. For recommendation results, the higher the better. For Simpson's paradox results, the lower the better. Positive improvements are bold and the highest is underlined.}
\vspace{-10pt}
\begin{tabular}{l|l|rrrr|rrrr|rrrr|rrrr}
\toprule
\multicolumn{2}{c}{\multirowcell{4}{Models}} & \multicolumn{8}{|c}{\textbf{ML100k}} & \multicolumn{8}{|c}{\textbf{Coat Shopping}}\\\cmidrule(lr){3-10}\cmidrule(lr){11-18}
\multicolumn{2}{c|}{} & \multicolumn{4}{c}{Recommendation} & \multicolumn{4}{c|}{Simpson's Paradox} & \multicolumn{4}{c}{Recommendation} & \multicolumn{4}{c}{Simpson's Paradox}\\\cmidrule(lr){3-10}\cmidrule(lr){11-18}
\multicolumn{2}{c|}{} & \multicolumn{2}{c}{nDCG@10} & \multicolumn{2}{c}{Hit@1} & \multicolumn{2}{c}{Gender} & \multicolumn{2}{c}{Age} & \multicolumn{2}{|c}{nDCG@10} & \multicolumn{2}{c}{Hit@1} & \multicolumn{2}{c}{Gender} & \multicolumn{2}{c}{Age} \\\cmidrule(lr){3-10}\cmidrule(lr){11-18}
\multicolumn{2}{c|}{} & value & imp & value & imp & value & imp & value & imp & value & imp & value & imp & value & imp & value & imp \\
\midrule
     \multirowcell{9}{MF}
        & Original & 0.3647 & - & 0.1490 & - & 0.1918 & - & 0.1864 & - & 0.2042 & - & 0.1561 & -  & 0.4415 & - & 0.3236 & - \\
        & $\text{CausE}$ & 0.3784 & \underline{\textbf{3.8\%}} & 0.1651 & \textbf{10.8\%} & 0.1205 & \textbf{37.2\%} & 0.1123 & \textbf{39.8\%} & 0.2100 & \textbf{2.8\%} & 0.1899 & \textbf{21.7\%} & 0.4462 & -1.1\% & 0.3124 & \textbf{3.5\%} \\
        & $\text{IPS}$ & 0.3696 & \textbf{1.3\%} & 0.1618 & \textbf{8.6\%} & 0.1966 & -2.5\% & 0.1954 & -4.8\% &  0.2261 & \textbf{10.7\%} & 0.2068 & \textbf{32.5\%} & 0.4392 &  \textbf{0.5\%} & 0.3217 & \textbf{0.6\%}\\
        & $\text{DCF}$ & 0.3686 & \textbf{1.1\%} & 0.1543 & \textbf{3.6\%} & 0.2190 & -14.2\% & 0.2091 & -12.2\% & 0.2209 & \textbf{8.2\%} & 0.1983 & \textbf{27.0\%} & 0.4615 & -4.5\% & 0.3622 & -11.9\% \\ 
        & $\text{DICE}$ & 0.3692 & \textbf{1.2\%} & 0.1543 & \textbf{3.6\%} & 0.2344 & -22.2\% & 0.2273 & -21.9\% & 0.2303 & \textbf{12.8\%} & 0.2068 & \textbf{32.4\%} & 0.5610 & -27.1\% & 0.4433 & -37.0\%\\
        & $\text{MACR}$ & 0.3671 & \textbf{0.7\%} & 0.1586 & \textbf{6.4\%} & 0.1516 & \textbf{21.0\%} & 0.1235 & \textbf{33.7\%} & 0.2197 & \textbf{7.6\%} & 0.2236 & \textbf{43.2\%} & 0.3941 & \textbf{10.7\%} & 0.2936 & \textbf{9.3\%}\\
        \cdashline{2-18}[.4pt/1pt]
        & $\text{CCF}_{\text{K1}}$ & 0.3661 & \textbf{0.4\%} & 0.1554 & \textbf{4.3\%} & 0.1834 & \textbf{4.4\%} & 0.1822 & \textbf{2.3\%} & 0.2324 & \underline{\textbf{13.8\%}} & 0.2363 & \underline{\textbf{51.4\%}} & 0.3547 & \underline{\textbf{19.7\%}} & 0.2621 & \underline{\textbf{19.0\%}}\\
        & $\text{CCF}_{\text{D1}}$ & 0.3781 & \textbf{3.7\%} & 0.1683 & \underline{\textbf{13.0\%}} & 0.1125 & \underline{\textbf{41.3\%}} & 0.1087 & \textbf{41.7\%} & 0.2061 & \textbf{0.9\%} & 0.2152 & \textbf{37.9\%} & 0.4217 & \textbf{4.5\%} & 0.3078 & \textbf{4.9\%}\\
        & $\text{CCF}_{\text{R1r}}$ & 0.3673 & \textbf{0.7\%} & 0.1554 & \textbf{4.3\%} & 0.1342 & \textbf{30.0\%} & 0.1195 & \textbf{35.9\%} & 0.2165 & \textbf{6.0\%} & 0.2194 & \textbf{40.6\%} & 0.3642 & \textbf{17.5\%} & 0.2810 & \textbf{13.2\%}\\
        & $\text{CCF}_{\text{R1n}}$ & 0.3734 & \textbf{2.4\%} & 0.1533 & \textbf{2.9\%} & 0.1217 & \textbf{36.5\%} & 0.1203 & \textbf{35.5\%} & 0.2089 & \textbf{2.3\%} & 0.2110 & \textbf{35.2\%} & 0.4012 & \textbf{9.1\%} & 0.2991 & \textbf{7.6\%}\\
        & $\text{CCF}_{\text{C}}$ & 0.3729 & \textbf{2.2\%} & 0.1597 & \textbf{7.2\%} & 0.1142 & \textbf{40.5\%} & 0.1062 & \underline{\textbf{43.0\%}} & 0.2150 & \textbf{5.3\%} & 0.2068 & \textbf{32.5\%} & 0.4206 & \textbf{4.7\%} & 0.3109 & \textbf{3.9\%}\\
        \midrule
        \multirow{9}{*}{GRU4Rec} & Original & 0.4087 & - & 0.1865 & - & 0.1460 & - & 0.1397 & - & 0.1147 & - & 0.0759 & - & 0.1252 & - & 0.1260 & -\\
        & $\text{CausE}$ & 0.4111 & \textbf{0.6\%} & 0.1908 & \textbf{2.3\%} & 0.1366 & \textbf{6.4\%} & 0.1313 & \textbf{6.0\%} & 0.1157 & \textbf{0.9\%} & 0.0802 & \textbf{5.7\%} & 0.1170  & \textbf{6.5\%} & 0.1006 & \textbf{20.2\%}\\
        & $\text{IPS}$ & 0.4136 & \textbf{1.2\%} & 0.1876 & \textbf{0.6\%} & 0.1480 & -1.4\% & 0.1292 & \textbf{7.5\%} & 0.1160 & \textbf{1.1\%} & 0.0802 & \textbf{5.7\%} & 0.1239 & \textbf{1.0\%} & 0.1204 & \textbf{4.4\%} \\
        & $\text{DCF}$ & 0.4158 & \textbf{1.7\%} & 0.1951 & \textbf{4.6\%} & 0.1392 & \textbf{4.7\%} & 0.1330 & \textbf{4.8\%}& 0.1174 & \textbf{2.4\%} & 0.0717 & -5.5\% & 0.1317 & -5.2\% & 0.1282 & -1.7\%\\
        & $\text{DICE}$ & 0.4158 & \textbf{1.7\%} & 0.1929 & \textbf{3.4\%} & 0.1325 & \textbf{9.2\%}& 0.1128 & \textbf{19.3\%}& 0.1273 & \textbf{11.0\%} & 0.0886 & \textbf{16.7\%} & 0.1362 & -8.8\% & 0.1101 & \textbf{12.6\%}\\
        & $\text{MACR}$ & 0.4211 & \textbf{3.0\%} & 0.1875 & \textbf{0.5\%} & 0.1304 & \textbf{10.7\%} & 0.1167 & \textbf{16.5\%} & 0.1289 & \textbf{12.8\%} & 0.0928 & \textbf{22.3\%} & 0.1437 & -14.8\% & 0.1521 & -20.7\%\\
        \cdashline{2-18}[.4pt/1pt]
        & $\text{CCF}_{\text{K1}}$ & 0.4225 & \textbf{3.4\%} & 0.2015 & \underline{\textbf{8.0\%}} & 0.1152 & \textbf{21.1\%} & 0.1088 & \underline{\textbf{22.1\%}} & 0.1170 & \textbf{2.0\%} & 0.0675 & -11.1\% & 0.1241 & \textbf{0.9\%} & 0.1225 & \textbf{2.8\%}\\
        & $\text{CCF}_{\text{D1}}$ & 0.4281 & \underline{\textbf{4.7\%}} & 0.1972 & \textbf{5.7\%} & 0.1235 & \textbf{15.4\%} & 0.1145 & \textbf{18.0\%} & 0.1226 & \textbf{6.9\%} & 0.0886 & \textbf{16.7\%} & 0.1185 & \textbf{5.4\%} & 0.1013 & \textbf{19.6\%}\\
        & $\text{CCF}_{\text{R1r}}$ & 0.4241 & \textbf{3.8\%} & 0.1972 & \textbf{5.7\%} & 0.1257 & \textbf{13.9\%} & 0.1174 & \textbf{16.0\%} & 0.1299 & \textbf{13.3\%} & 0.0802 & \textbf{5.7\%} & 0.1139 & \textbf{9.0\%} & 0.1093 & \textbf{13.3\%}\\
        & $\text{CCF}_{\text{R1n}}$ & 0.4235 & \textbf{3.6\%} & 0.2015 & \underline{\textbf{8.0\%}} & 0.1143 & \underline{\textbf{21.7\%}} & 0.1113 & \textbf{20.3\%} & 0.1264 & \textbf{10.2\%} & 0.0802 & \textbf{5.7\%}  & 0.1082 & \textbf{13.6\%} & 0.1154 & \textbf{8.4\%}\\
        & $\text{CCF}_{\text{C}}$ & 0.4238 & \textbf{3.7\%} & 0.2015 & \underline{\textbf{8.0\%}} & 0.1245 & \textbf{14.7\%} & 0.1095 & \textbf{21.6\%} & 0.1352 & \underline{\textbf{17.9\%}} & 0.0970 & \underline{\textbf{27.8\%}} & 0.1047 & \underline{\textbf{16.4\%}} & 0.0895 & \underline{\textbf{29.0\%}}\\
        \midrule
        \multirow{9}{*}{NCR} & Original & 0.4227 & - & 0.1972 & - & 0.1054 & - & 0.0828 & - & 0.2608 & - & 0.0506 & - & 0.1320 & - & 0.1128 & - \\
        & $\text{CausE}$ & 0.4234 & \textbf{0.2\%} & 0.2090 & \textbf{6.0\%} & 0.0829 & \underline{\textbf{21.3\%}}  & 0.0798 & \textbf{3.6\%} & 0.2813 & \textbf{7.9\%} & 0.0886 & \textbf{75.1\%} & 0.1236 & \textbf{6.4\%} & 0.1057 & \textbf{6.3\%}\\
        & $\text{IPS}$ & 0.4237 & \textbf{0.2\%} & 0.2036 & \textbf{3.2\%} & 0.0874 & \textbf{17.1\%} & 0.0826 & \textbf{0.2\%} & 0.2916 & \textbf{11.8\%} & 0.1350 & \textbf{166.8\%}  & 0.1285 & \textbf{2.7\%} & 0.1094 & \textbf{3.0\%}\\
        & $\text{DCF}$ & 0.4201 & -0.6\% & 0.1940 & -1.6\% & 0.0859 & \textbf{18.5\%}& 0.0789 & \textbf{4.7\%} & 0.2689 & \textbf{3.1\%} & 0.0802 & \textbf{58.5\%} & 0.1347 & -2.0\% & 0.1135 & -0.6\%\\
        & $\text{DICE}$ & 0.4199 & -0.7\% & 0.1994 & \textbf{1.1\%} & 0.1211 & -14.9\% & 0.1149 & -38.8\% & 0.2892 & \textbf{10.9\%} & 0.0928 & \textbf{83.4\%} & 0.1421 & -7.7\% & 0.1207 & -7.0\%\\
        & $\text{MACR}$ & 0.4231 & \textbf{0.1\%} & 0.2101 & \textbf{6.5\%} & 0.0892 & \textbf{15.4\%} & 0.0818 & \textbf{1.2\%} & 0.3011 & \textbf{15.5\%} & 0.1392 & \underline{\textbf{175.1\%}} & 0.1175 & \textbf{11.0\%} & 0.0987 & \textbf{12.5\%}\\
        \cdashline{2-18}[.4pt/1pt]
        & $\text{CCF}_{\text{K1}}$ & 0.4094 & -3.1\% & 0.1940 & -1.6\% & 0.1047 & \textbf{0.7\%} & 0.0835 & -0.8\% & 0.2896 & \textbf{11.0\%} & 0.1139 & \textbf{125.1\%} & 0.1213 & \textbf{8.1\%} & 0.1083 & \textbf{4.0\%}\\
        & $\text{CCF}_{\text{D1}}$ & 0.4144 & -2.0\% & 0.1897 & -3.8\% & 0.1103 & -4.6\% & 0.0844 & -1.9\% & 0.3098 & \underline{\textbf{18.8\%}} & 0.1308 & \textbf{158.5\%} & 0.1139 & \underline{\textbf{13.7\%}} & 0.1021& \textbf{9.5\%}\\
        & $\text{CCF}_{\text{R1r}}$ & 0.4271 & \textbf{1.0\%} & 0.2004 & \textbf{1.6\%} & 0.0854 & \textbf{19.0\%} & 0.0713 & \underline{\textbf{13.9\%}} & 0.2874 & \textbf{10.2\%} & 0.1013 & \textbf{100.2\%} & 0.1229 & \textbf{6.9\%} & 0.1102 & \textbf{2.3\%}\\
        & $\text{CCF}_{\text{R1n}}$ & 0.4195 & -0.8\% & 0.2111 & \underline{\textbf{7.0\%}} & 0.0881 & \textbf{16.4\%} & 0.0803 & \textbf{3.0\%} & 0.2816 & \textbf{8.0\%} & 0.0970 & \textbf{91.7\%} & 0.1243& \textbf{5.8\%} & 0.1079 & \textbf{4.3\%}\\
        & $\text{CCF}_{\text{C}}$ & 0.4274 & \underline{\textbf{1.1\%}} & 0.2058 & \textbf{4.4\%} & 0.0871 & \textbf{17.4\%} & 0.0733 & \textbf{11.5\%} & 0.3095 & \textbf{18.7\%} & 0.1266 & \textbf{150.2\%} & 0.1188 & \textbf{10.0\%} & 0.0968 & \underline{\textbf{14.2\%}}\\
        \bottomrule
\end{tabular}
    \label{tab:performance}
    \vspace{-10pt}
\end{table*}

\vspace{-5pt}
\subsection{Data Description}
Our experiments are conducted on two types of datasets. The first type is frequently used benchmark dataset \textbf{MovieLen-100k}\footnote{\url{https://grouplens.org/datasets/movielens/}}. 
For the second type, to better show that our framework can help to capture users' preference, we apply our framework on the \textbf{Coat Shopping}\footnote{\url{https://www.cs.cornell.edu/~schnabts/mnar/}} dataset. A special property of this dataset is that the testing data are collected from randomized trials, i.e., users give feedback on random items.

\subsection{Baseline Models}
We employ five causal frameworks for comparison. \textbf{CausE}~\cite{bonner2018causal} is a direct intervention model, which creates randomized treatment data for causal learning. 
\textbf{IPS}~\cite{saito2020unbiased} is an Inverse Propensity Scoring-based model, which uses a user-independent propensity estimator to re-weight the training samples. \textbf{DCF}~\cite{wang2020causal} is a deconfounded recommender, which uses an exposure model to construct a substitute confounder.
\textbf{DICE}~\cite{zheng2021disentangling} is a framework for disentangling user interest and conformity for recommendation with causal embedding. \textbf{MACR}~\cite{wei2021model} is a model-agnostic framework for alleviating popularity bias issue in recommender systems.

Meanwhile, we test five versions of our framework. $\textbf{CCF}_{\text{K1}}$, $\textbf{CCF}_{\text{D1}}$, $\textbf{CCF}_{\text{R1r}}$, $\textbf{CCF}_{\text{R1n}}$ are CCF in discrete space under different heuristic rules.
$\textbf{CCF}_{\text{C}}$ is CCF in continuous space.

We apply all above frameworks on three base recommendation models, including a matching model (MF), a sequential model (GRU4Rec) and a reasoning model (NCR). \textbf{MF}~\cite{rendle2012bpr} uses Matrix Factorization \cite{koren2009matrix} as the prediction function under Bayesian personalized ranking.
\textbf{GRU4Rec}~\cite{hidasi2016session} uses Gated Recurrent Units (GRU) to capture sequential patterns.
\textbf{NCR}~\cite{chen2020neural} organizes the logic expressions as neural networks for reasoning and recommendation.

\begin{figure*}[t!]
\captionsetup[sub]{font=small,labelfont=normalfont,textfont=normalfont}
    \centering
    \begin{subfigure}{0.24\textwidth}
        \includegraphics[scale=0.35]{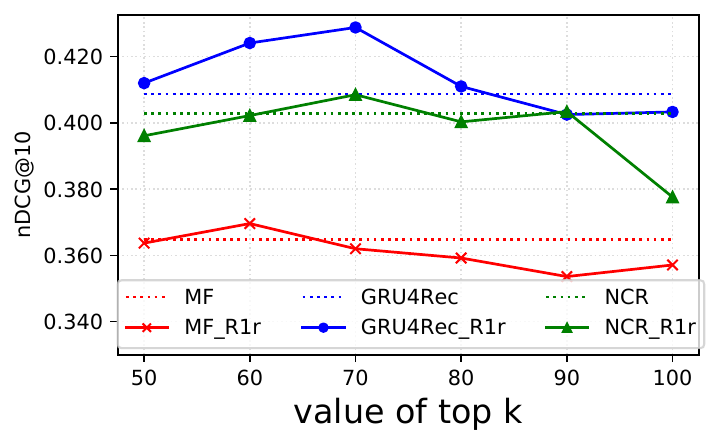}
        \vspace{-15pt}
        \caption{Discrete Recommendation}
    \end{subfigure}
    \begin{subfigure}{0.24\textwidth}
        \includegraphics[scale=0.35]{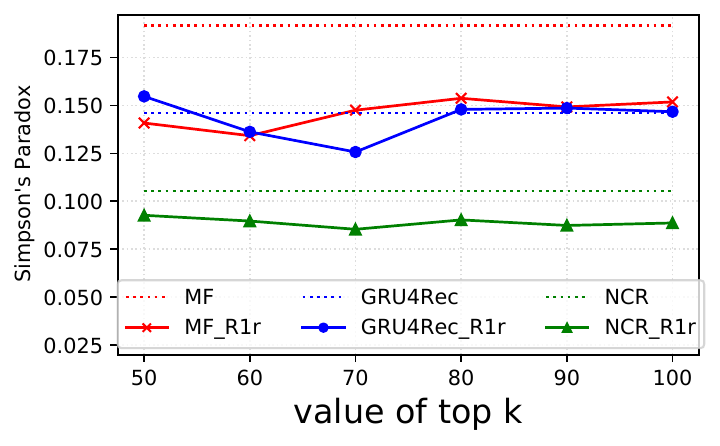}
        \vspace{-15pt}
        \caption{Discrete Simpson's Paradox}
    \end{subfigure}
    \begin{subfigure}{0.24\textwidth}
        \includegraphics[scale=0.35]{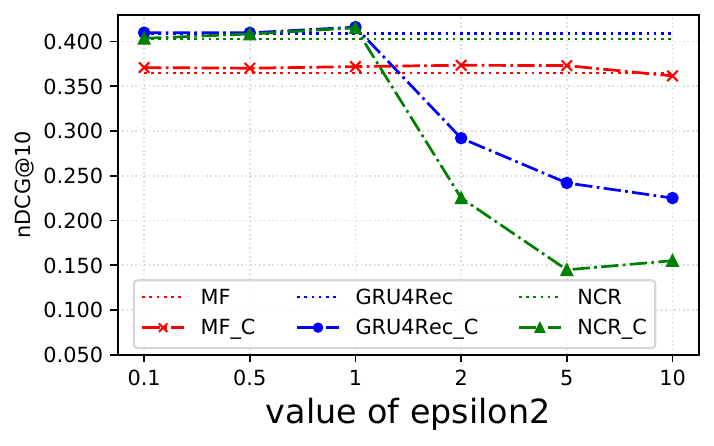}
        \vspace{-15pt}
        \caption{Continuous Recommendation}
    \end{subfigure}
    \begin{subfigure}{0.24\textwidth}
        \includegraphics[scale=0.35]{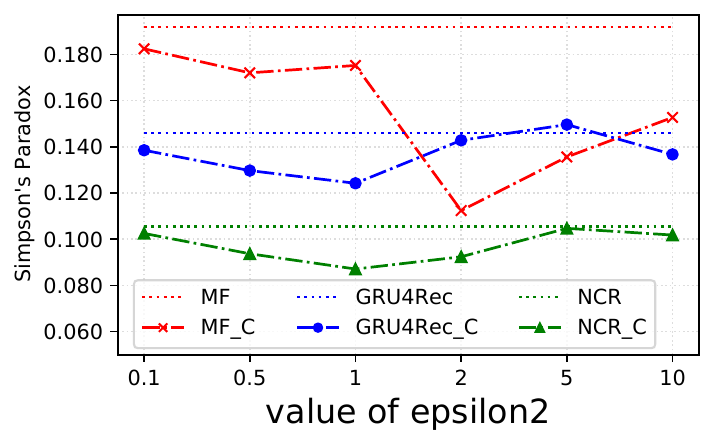}
        \vspace{-15pt}
        \caption{Continuous Simpson's Paradox}
    \end{subfigure}
    \vspace{-10pt}
    \caption{Recommendation results (nDCG@10) and Simpson's paradox results (grouped by gender) on ML100k with different counterfactual selection parameters. (a) and (b) are discrete versions with R1r heuristic rule under parameter $k$. (c) and (d) are continuous versions under parameter $\epsilon_2$.}
    \label{fig:counterfactual_selection}
    \vspace{-10pt}
\end{figure*}

\begin{figure*}[t!]
\captionsetup[sub]{font=small,labelfont=normalfont,textfont=normalfont}
    \centering
    \begin{subfigure}{0.24\textwidth}
        \includegraphics[scale=0.35]{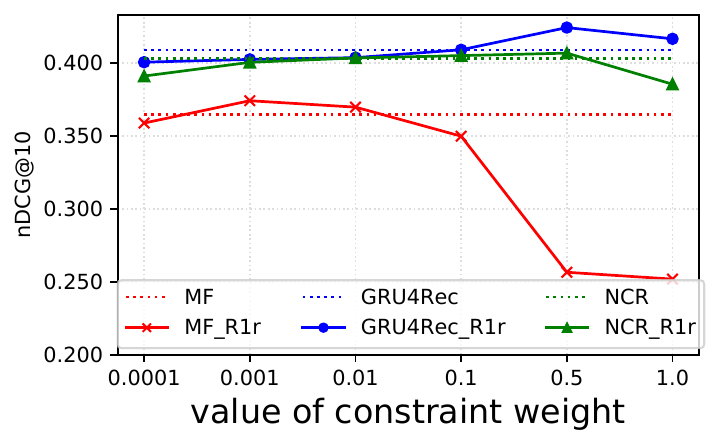}
        \vspace{-15pt}
        \caption{Discrete Recommendation}
    \end{subfigure}
    \begin{subfigure}{0.24\textwidth}
        \includegraphics[scale=0.35]{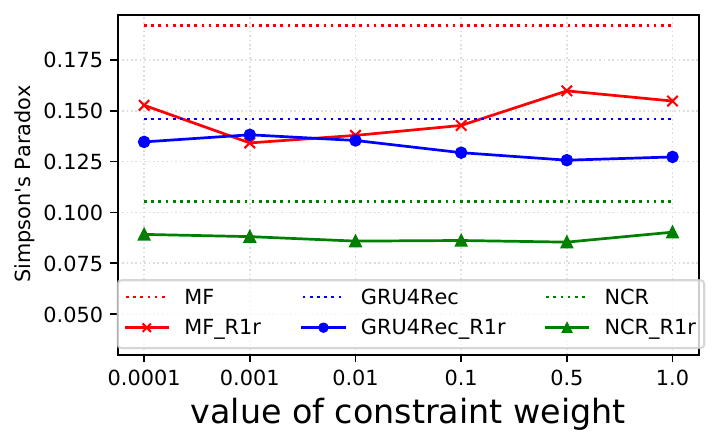}
        \vspace{-15pt}
        \caption{Discrete Simpson's Paradox}
    \end{subfigure}
    \begin{subfigure}{0.24\textwidth}
        \includegraphics[scale=0.35]{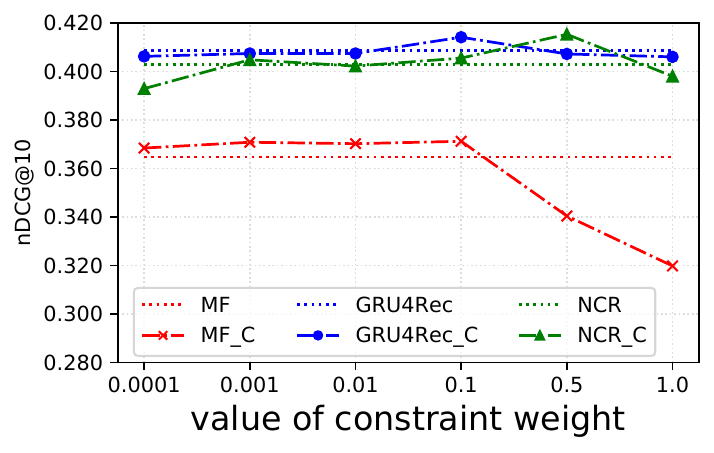}
        \vspace{-15pt}
        \caption{Continuous Recommendation}
    \end{subfigure}
    \begin{subfigure}{0.24\textwidth}
        \includegraphics[scale=0.35]{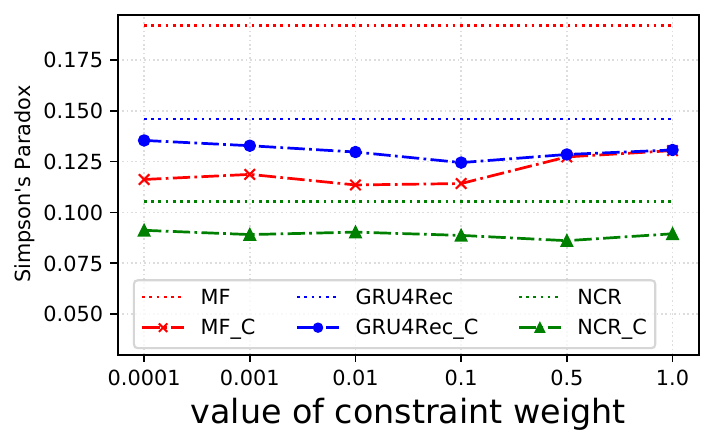}
        \vspace{-15pt}
        \caption{Continuous Simpson's Paradox}
    \end{subfigure}
    \vspace{-10pt}
    \caption{Recommendation results (nDCG@10) and Simpson's paradox results (grouped by gender) on ML100k with different counterfactual constraint weight $\omega$. (a) and (b) are discrete versions with R1r heuristic rule. (c) and (d) are continuous versions.}
    \label{fig:constraint_weight}
    \vspace{-10pt}
\end{figure*}

\begin{figure*}[t!]
\captionsetup[sub]{font=small,labelfont=normalfont,textfont=normalfont}
    \centering
    \begin{subfigure}{0.24\textwidth}
        \includegraphics[scale=0.35]{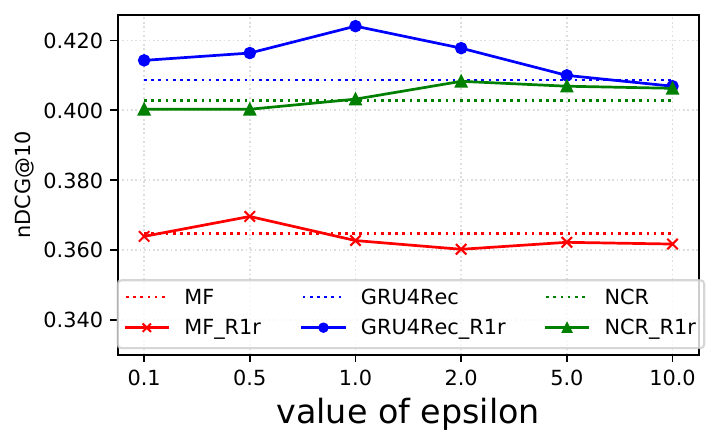}
        \vspace{-15pt}
        \caption{Discrete Recommendation}
    \end{subfigure}
    \begin{subfigure}{0.24\textwidth}
        \includegraphics[scale=0.35]{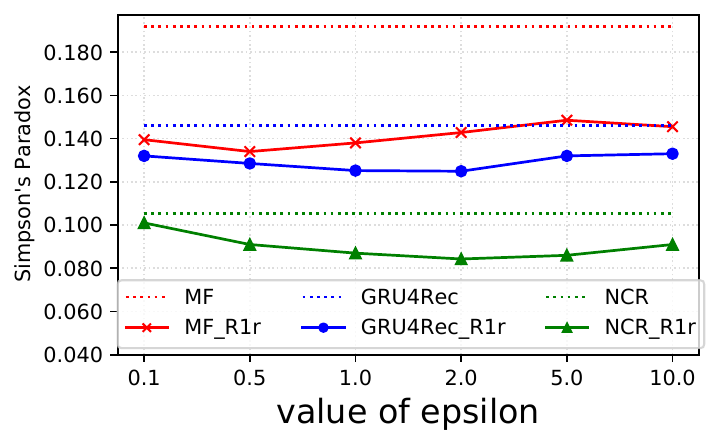}
        \vspace{-15pt}
        \caption{Discrete Simpson's Paradox}
    \end{subfigure}
    \begin{subfigure}{0.24\textwidth}
        \includegraphics[scale=0.35]{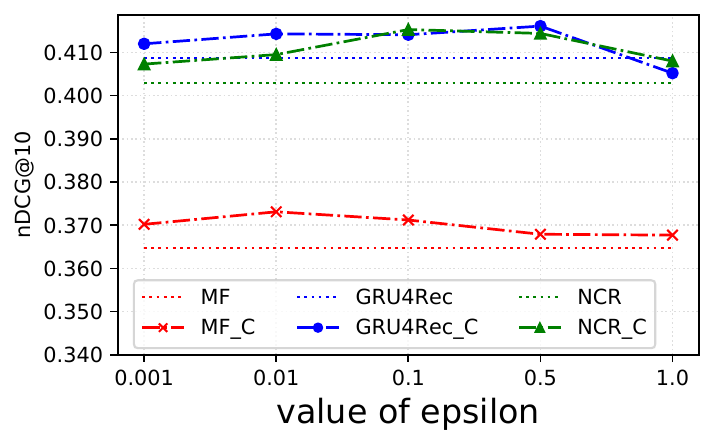}
        \vspace{-15pt}
        \caption{Continuous Recommendation}
    \end{subfigure}
    \begin{subfigure}{0.24\textwidth}
        \includegraphics[scale=0.35]{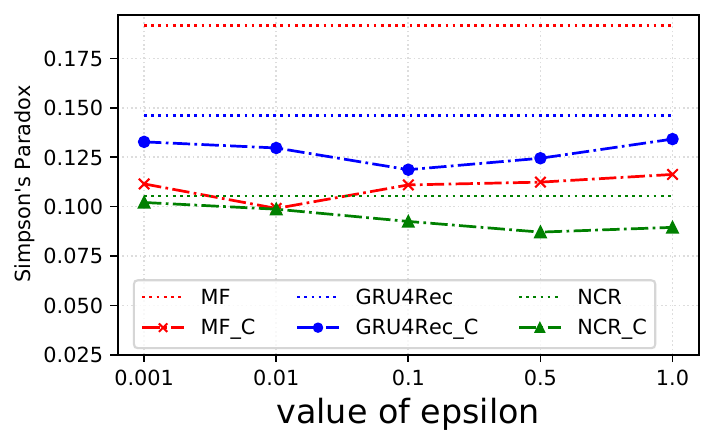}
        \vspace{-15pt}
        \caption{Continuous Simpson's Paradox}
    \end{subfigure}
    \vspace{-10pt}
    \caption{Recommendation results (nDCG@10) and Simpson's paradox results (grouped by gender) on ML100k with different counterfactual constraint threshold $\epsilon$ ($\epsilon_1$ for continuous version). (a) and (b) are discrete versions with R1r heuristic rule. (c) and (d) are continuous versions.}
    \label{fig:epsilon}
    \vspace{-15pt}
\end{figure*}

\vspace{-1ex}
\subsection{Overall Performance}

We answer \textbf{RQ1} and \textbf{RQ2} in this section by showing the recommendation performance and Simpson's paradox performance of applying causal frameworks (CausE, IPS, DCF, DICE, MACR, $\text{CCF}_*$) on the three recommendation models in Table \ref{tab:performance}. 

For recommendation performance, we can see that in most cases causal frameworks can bring positive improvement to the recommendation models. Comparing all frameworks, we see that for all of the recommendation models, the largest average improvement over four datasets is mostly brought by our CCF framework.

For Simpson's paradox evaluation, we follow the paradox detection method in \cite{jadidinejad2021simpson}.
More specifically, we split users into two groups according to gender (a binary feature in the dataset) or age (the split threshold is 35). Similar to the example in Figure \ref{fig:paradox}(b), each user only recommends his or her top-$K$ items of highest predicted scores, where $K$ is set to 50. We randomly sample 100 users for each item and calculate the percentage of item pairs that have Simpson's paradox. 
The Simpson's paradox performance is shown in Table \ref{tab:performance}, and more results of Simpson's paradox mitigation are provided in Section \ref{sec:paradox}. 
We can see that our CCF framework mitigates Simpson's paradox in most cases while improving recommendation performance.
In contrast, other causal frameworks may improve the recommendation performance but not necessarily mitigate the Simpson's paradox since they are designed through other perspectives. 
We calculate the Pearson correlation coefficient between recommendation improvement (average of improvements on nDCG@10 and Hit@1) and Simpson's paradox mitigation (average of improvements on gender and age) in Table \ref{tab:pearson}. The positive correlation indicates that mitigating Simpson's paradox will help improve recommendation performance.

\begin{table}[]
    \centering
    \small
    \begin{tabular}{ccc}
    \toprule
         & \textbf{ML100k} & \textbf{Coat Shopping} \\\midrule
        MF & 0.52 & 0.33 \\
        GRU4Rec & 0.84 & 0.18\\
        NCR & 0.43 & 0.67\\
        \bottomrule
    \end{tabular}
    \caption{
    Pearson correlation coefficient to measure the relationship between recommendation improvement and Simpson's paradox mitigation.}
    \vspace{-25pt}
    \label{tab:pearson}
\end{table}

As we mentioned before, CausE improves performance by splitting the observational training data into approximately randomized data, IPS improves performance by re-weighting the observational training data, DCF improves performance by reconstructing a substitute confounder, DICE improves performance by adopting separate embeddings for interest and conformity to disentangle them, and MACR improves performance by eliminating the popularity bias through removing the direct effect between item properties and the ranking score. 
All these frameworks only consider the real-world examples though with different techniques, however, the CCF framework not only considers real-world examples but also involves counterfactual examples, which helps to mitigate the Simpson's paradox for making better decision and improving the recommendation performance.

\vspace{-2ex}
\subsection{Analyzing Counterfactual Examples}
In this section, we aim to answer research questions \textbf{RQ3} and \textbf{RQ4}. 
We first dig into the difference between different heuristic rules.
We then show the necessity of the selection process after generation.

\subsubsection{\textbf{Difference between Heuristic Rules}}
In this section, we focus on the discrete versions of CCF and discuss the effect of different heuristic rules.
Among the heuristic rules in Table \ref{tab:heuristic_rules}, K1 and D1 generate much fewer counterfactual histories than R1, because K1 and D1 are limited by the number of interactions in the user's real history, while R1 can replace each interacted item with a large number of possible items.
As a result, it is more difficult for K1 and D1 to get satisfied counterfactual examples in the selection process when $k$ is small, where $k$ is the top-$k$ selection threshold introduced in Section \ref{sec:select}. 
Considering the difficulty of generating qualified counterfactual examples, the R1 rules 
are intuitively better than D1 and K1.
This is consistent with the experimental results.

\begin{figure}[t]
    \centering
    \includegraphics[width=\linewidth]{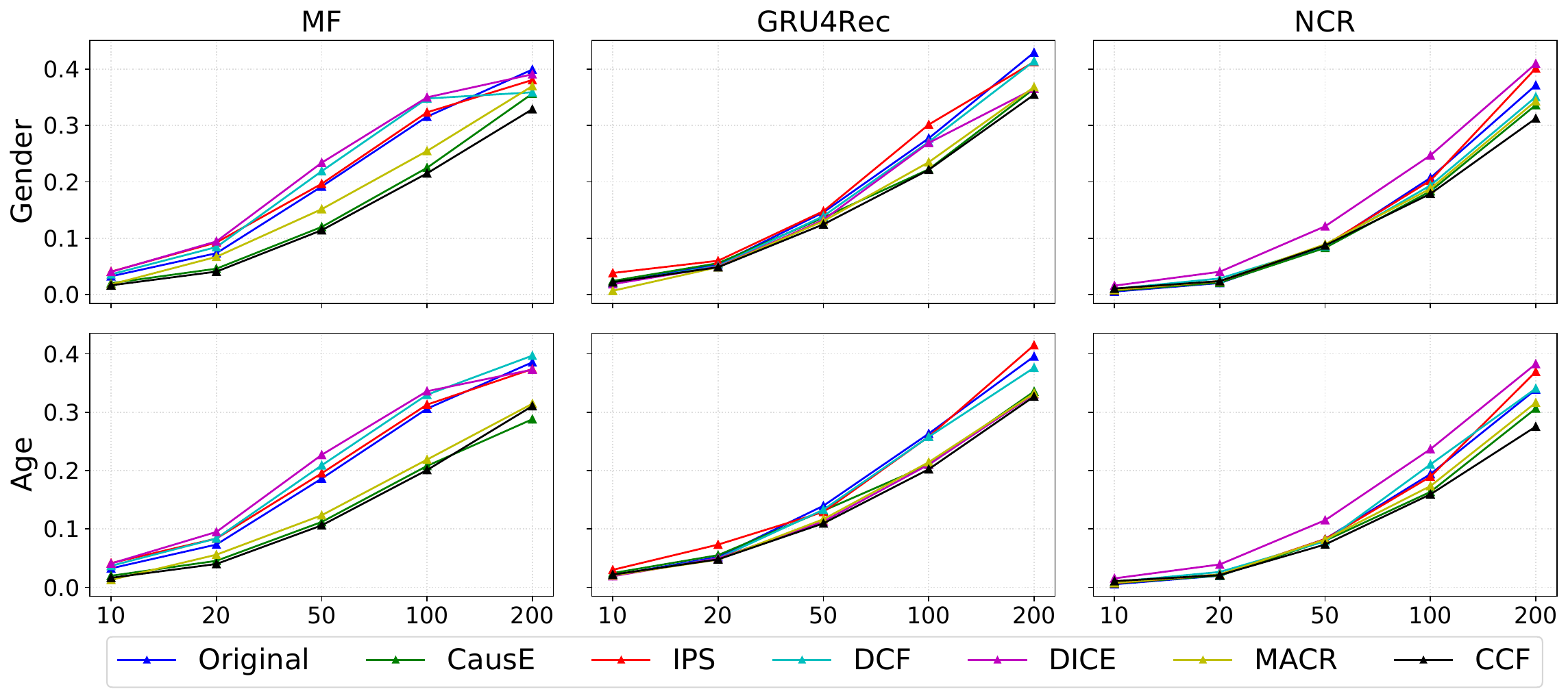}
    \vspace{-20pt}
    \caption{The percentage of item pairs showing paradox when grouping by gender (above) and age (below) on ML100k. 
    $x$-axis is the number of items recommended by each user.}
    \vspace{-15pt}
    \label{fig:paradox_results}
\end{figure}

\subsubsection{\textbf{Counterfactual Example Selection}}
The selection is based on parameter $k$ (Section \ref{sec:select}) and $\epsilon_2$ in Eq.\eqref{eq:continuous_learning} for CCF discrete and continuous version, respectively.
We first examine the discrete versions. We plot the recommendation performance on nDCG@10 and the Simpson's paradox results under different $k$ on Movielens-100k in Figure \ref{fig:counterfactual_selection}(a-b), where users are grouped by gender and the R1r rule is used. Other rules and datasets have similar observations.
We see that when $k$ is properly chosen, our framework will mitigate Simpson's paradox and improve the performance.
However, when $k$ is too large---such as $k=100$ so that all the generated counterfactual examples are selected---the counterfactual constraint will mislead the causal preference estimation and lead to relatively more paradox thus hurt the performance. This observation is consistent with the theory of conditional intervention (Section \ref{sec:conditional_intervention}). 

For continuous version, we plot recommendation performance and Simpson's paradox results under different $\epsilon_2$ in Figure \ref{fig:counterfactual_selection}(c-d). When $\epsilon_2$ is small, the counterfactual embedding $\mathbf{x}^\prime$ is very close to the real embedding $\mathbf{x}$ (Eq.\eqref{eq:continuous}\eqref{eq:continuous_learning}), therefore, the estimation of preference after applying CCF has no much difference from the original preference, because the counterfactual constraint in Eq.\eqref{eq:continuous}\eqref{eq:continuous_learning} is easily satisfied. In contrast, if $\epsilon_2$ is too large, $\mathbf{x}^\prime$ will be too far away from the real embedding $\mathbf{x}$, and if we force their predictions to be close, causal preference will not be correctly estimated thus the performance will decrease. 

\vspace{-1ex}
\subsection{Analyzing Counterfactual Constraints}
There are two important parameters for the counterfactual constraint---parameter $\omega$ in Eq.\eqref{eq:discrete_learning} and \eqref{eq:continuous_learning},
and parameter $\epsilon$ in Eq.\eqref{eq:discrete_learning} (or $\epsilon_1$ in Eq.\eqref{eq:continuous_learning}).
In this section, we provide the answers to \textbf{RQ5}. We will discuss the two parameters separately in the following.

\subsubsection{\textbf{Counterfactual Constraint Weight}}

Given loss function as Eq.\eqref{eq:discrete_learning} and \eqref{eq:continuous_learning}, the larger the counterfactual constraint weight $\omega$, the more likely the results will follow the constraint.  
We tune the $\omega$ while keeping other parameters fixed. The results of nDCG@10 are shown in Figure \ref{fig:constraint_weight}(a)(c). The results of Simpson's paradox (grouped by gender) are provided in Figure \ref{fig:constraint_weight}(b)(d).
We see that in most cases the performance would first getting better and then worse, meaning that the constraint is useful for recommendation and Simpson's paradox mitigation but it also requires a good balance with the original loss. When $\omega$ is too small, the constraint has little effect on the total loss, leading to only slight improvement or even slight decrease considering the larger model complexity. In contrast, if $\omega$ is too large, the constraint loss will dominate the total loss, and thus the recommendation performance is significantly decreased since the original loss does not take too much effect. 
Meanwhile, in this case, the causal preference may not be accurately estimated thus hurt Simpson's paradox mitigation.
Overall, the weight needs to be carefully specified in practice, and compared with an overly large weight, a relatively smaller weight would be preferred.

\subsubsection{\textbf{Counterfactual Constraint Threshold}}

The counterfactual constraint threshold (i.e. $\epsilon$ in Eq.\eqref{eq:discrete} and $\epsilon_1$ in Eq.\eqref{eq:continuous}) controls how rigorous the constraint is.
We plot nDCG@10 with different threshold in Figure \ref{fig:epsilon}. We see that the performance first getting better and then worse and finally tend to be flat when the threshold is large enough. When the threshold is too small, the constraint would be too tight and it makes the model less capable of handling the potential errors in counterfactual examples.
When the threshold is too large, we are actually applying no constraint, since the difference between real and counterfactual examples' prediction would always be smaller than the threshold, and thus the $L_c$ in Eq.\eqref{eq:discrete_learning} and \eqref{eq:continuous_learning} would be 0 in most cases.
As a result, the performance becomes relatively flat when the threshold is large enough.

\subsection{Influence of Recommendation Length on Simpson's Paradox Mitigation}\label{sec:paradox}
Figure \ref{fig:paradox_results} shows the percentage of item pairs that have paradox, with different $K$ ranging from 10 to 200 and users are grouped by gender or age. The figure shows three base recommendation models under all frameworks, where CCF is the continuous version. Coat dataset has similar observations. 
From the results, we can see that CCF framework is able to reduce paradox compared with the original model in the most cases. Additionally, CCF (i.e., the black line in Figure \ref{fig:paradox_results}) is almost always the lowest line in all sub-figures, showing that CCF achieves the best performance compared with other frameworks in terms of mitigating paradox.

\section{Conclusions and Future Work}\label{sec:conclusion}

In this paper, we proposed a causal framework (CCF) for mitigating Simpson's Paradox in recommendation.
We provided a conditional intervention approach to estimating the $P(y|u,do(v))$ and proposed a flexible counterfactual constrained learning framework which is applicable to many recommendation models.
Experiments show that CCF helps to mitigate Simpson's paradox and improve the performance of the matching-, sequential- and reasoning-based models.
The CCF framework is flexible and can be extended in various dimensions in the future, such as extending the causal graph to more complicated graphs for more complex recommendation scenarios. 
We will explore these possibilities in the future.

\section*{Acknowledgment}
This work was supported in part by NSF IIS-1910154,  IIS-2007907, IIS-2046457 and IIS-2127918. Any opinions, findings, conclusions or recommendations expressed in this material are those of the authors and do not necessarily reflect those of the sponsors.

\clearpage
\bibliographystyle{ACM-Reference-Format}
\balance
\bibliography{reference}

\end{document}